\documentclass[12pt]{article}
\usepackage{amsmath}
\usepackage{amsfonts}
\usepackage{amssymb}
\usepackage{graphicx}
\usepackage{color}

\numberwithin{equation}{section}

\makeatletter
\newcommand{\rd}{\@ifnextchar^{\DIfF}{\DIfF^{}}}
\def\DIfF^#1{%
   \mathop{\mathrm{\mathstrut d}}%
   \nolimits^{#1}\gobblespace}
\def\gobblespace{\futurelet\diffarg\opspace}
\def\opspace{%
   \let\DiffSpace\!%
   \ifx\diffarg(%
   \let\DiffSpace\relax
   \else
   \ifx\diffarg[%
   \let\DiffSpace\relax
   \else
   \ifx\diffarg\{%
   \let\DiffSpace\relax
   \fi\fi\fi\DiffSpace}
\providecommand*{\dder}[3][]{%
\frac{\rd^{#1}#2}{\rd #3^{#1}}}
\providecommand*{\pder}[3][]{%
\frac{\partial^{#1}#2}{\partial #3^{#1}}}
\providecommand*{\iu}%
{\ensuremath{\mathrm{i}\,}}

\def\d{\delta}

\def\rf#1{(\ref{eq:#1})}
\def\lab#1{\label{eq:#1}}

\begin{document}


   \begin{center}
 \Large\textbf{A Symmetric System of Mixed Painlev\'e III - V Equations
 and its Integrable Origin
}
\end{center}
\vskip 9 mm
\begin{center}
\begin{minipage}[t]{70mm}
{\bf H. Aratyn}\\
Department of Physics,\\
University of Illinois at Chicago,\\
845 W. Taylor St.,\\
Chicago, IL 60607\\
\end{minipage}
\begin{minipage}[t]{70mm}
{\bf J. F. Gomes, 
D. V. Ruy and A. H. Zimerman}\\
Instituto de F\'isica Te\'orica-UNESP\\
Rua Dr Bento Teobaldo Ferraz 271, Bloco II\\
01140-070, S\~ao Paulo, Brazil\\
\end{minipage}
\\
\end{center}
\vskip 9 mm
\abstract{A  mixed symmetric  Painlev\'e III - V model 
which describes a  hybrid of both equations is defined and 
obtained by successive
self-similarity and Dirac Lagrange multiplier reductions from
an integrable $4$-boson hierarchy.
}

\section{Introduction}
This paper deals with a system of nonlinear ordinary differential 
equations  containing both Painlev\'e III and V equations as well as
a new equation that passes the  basic Painlev\'e test
and possesses invariance under $A^{(1)}_1$ extended affine Weyl group.
This system of equations emerges through 
self-similarity 
and Dirac reductions performed on a 
special class of integrable models \cite{ANPZ,agz} referred to as
multi-boson Lax hierarchies and presented here in a Hamiltonian 
setting.

As postulated by Ablowitz, Ramani and Segur (ARS) \cite{ARS} 
the partial differential evolution 
equations of integrable hierarchies reduce in
self-similarity limit 
to ordinary differential equations with solutions that have no movable
critical points other than poles. 
This feature is known as Painlev\'e property.

The self-similarity reduction 
applied to  multi-boson Lax hierarchies \cite{ANPZ,agz}  leads to 
higher Painlev\'e equations invariant under 
extended affine Weyl groups $A^{(1)}_{2m}$ or $A^{(1)}_{2m-1}, m=1,2 {\ldots} $  
\cite{NoumiYamada98a,NoumiYamada98b,Noumiwkb,noumibk}.
More specifically, the $4$-boson model considered in this paper 
reduces in self-similarity limit 
to $A^{(1)}_4$ Painlev\'e  equations.
In another well-known
example the $2$-boson model reduces 
to $A^{(1)}_2$ Painlev\'e IV equations (see e.g. \cite{agz:2010}).
A construction given previously 
in \cite{agz,agz:2010,AGZ:2011}
involved further reduction of $A^{(1)}_4$ Painlev\'e  equation
to $A^{(1)}_3$  Painlev\'e V equation. 

In this paper  we are able to address the following question. How to reduce 
integrable models of $2m$-boson type to Painlev\'e equations
with symmetry structures other than  $A^{(1)}_{2m}$ or $A^{(1)}_{2m-1}$ 
($m=2$ in this paper)? For instance the Painlev\'e III equation
with $B^{(1)}_2$ extended affine Weyl symmetry is here obtained as
a limit of Dirac reduction of the  $A^{(1)}_{4}$ Hamiltonian system.
The two-step reduction of the $4$-boson model, first 
to $A^{(1)}_{4}$ equations by  the self-similarity reduction then followed 
by the Dirac reduction involving Lagrange multipliers leads
to  
a new system of mixed symmetric Painlev\'e III - V equations
that embed $A^{(1)}_3$ Painlev\'e V equation and $B^{(1)}_2$
Painlev\'e {III} equation. 
Explicitly this new system is governed by the following equations:
\begin{align}
z f_{i,\,z} &=f_i  f_{i+2} \big(
 f_{i+1}- f_{i+3} 
 \big)+(-1)^{i} f_i \,\big(\alpha_{1}+\alpha_{3}+C\big)  \lab{big}
\\ 
&+
\alpha_i\big(f_{i}+f_{i+2}\big)
-(-1)^{[i/2]} \epsilon_{i+1} \big(f_{i+1}+f_{i+3}\big), \qquad i=0,1,2,3
\nonumber
\end{align}
where $f_i=f_{i+4}$ and the symbol $[i/2]$ is $i/2$, if $i$ is even or $(i+1)/2$, 
if $i$ is odd. Equations \rf{big} have two features that distinguish
them from standard {symmetric} 
$A^{(1)}_{3}$ Painlev\'e V equations
\cite{NoumiYamada98a,NoumiYamada98b,MOK01}. 
First, one notices presence of additional terms that 
contain ``deformation parameters'' $\epsilon_i$. These parameters
satisfy periodicity conditions $\epsilon_i=\epsilon_{i+2}$
that reduce their number to two: $\epsilon_0$  
and $\epsilon_1$. Secondly, in addition to usual parameters $\alpha_i, i=0,1,2,3$ of the
affine Weyl structure the model also depends explicitly on an arbitrary 
parameter $C$. 

Equations \rf{big} possess two
integration constants $r_0,r_1$ appearing in relations:
\begin{equation}
f_1+f_3 =r_1\, z^{-C}, \quad
f_0+f_2 = r_0\, z^{(C+\Omega) } \, ,
\lab{condcbig}
\end{equation}
where 
\begin{equation}
\Omega= \alpha_0+\alpha_1+\alpha_2+\alpha_3 \ne 0\, .
\lab{omegadef}
\end{equation}
For two special values of {$C$}: $C=0$ or $C=-\Omega$ 
the combinations $f_1+f_3$ or $f_0+f_2$ become equal to integration
constants $r_1$ or $r_0$, respectively.

For $C$ satisfying the condition
\begin{equation}
 C=- \Omega/2 
\,.\lab{Calphas}
\end{equation}
the terms $f_1+f_3$ and $f_0+f_2$ are proportional to each other by a constant.
Such condition was assumed in \cite{NoumiYamada98a,NoumiYamada98b,MOK01}.
When {the} condition \rf{Calphas} holds 
equations \rf{big} describe either Painlev\'e  III or Painlev\'e V equation.
More explicitly,  equations \rf{big} will turn into Painlev\'e  III or 
Painlev\'e V equations
depending on whether the integration constant $r_1$ vanishes or has a
non-zero value (to simplify the argument we assumed here that 
constants  $r_0$ and $\epsilon_0$ remain different from zero).
Thus in the case of \rf{Calphas} (and $r_0 \ne 0$)
equations \rf{big} have   effectively only one
deformation parameter $r_1$ determining transition between Painlev\'e III
and V cases.

The symmetry structure of equations \rf{big} possesses additional features when $C$ fails to satisfy the 
condition  \rf{Calphas}, which we illustrate here 
by considering $C=0$. In this case (and with
$r_0=1, \epsilon_1=0$) this model is described
by a second order differential equation:
\begin{equation} \begin{split}
q_{zz} & = - \frac1z q_z+ \left( \frac{1}{2q}+\frac{1}{2(q-r_1)} \right)
(q_z^2 -\epsilon_0^2)
- \frac{(\alpha r_1 -\epsilon_0 z \alpha_3) q}{z^2 (q-r_1)} \\
&-\frac{\beta r_1 (q-r_1)+\epsilon_0 r_1 z \alpha_1 }{z^2 q} 
- \frac{\gamma q (q-r_1) - \epsilon_0  (1+\alpha_1)}{z}
+\frac12 q (q-r_1) (2q-r_1)\, ,
\lab{sP5q}
\end{split}
\end{equation}
where $q=f_1$ and with $\alpha, \beta, \gamma $ parameters 
to be defined below in relation \rf{abc}. The other 
two parameters present in the above equation are: 
a deformation parameter  $\epsilon_0$   
and an integration constant $r_1$.
For non-zero values of these two parameters
equation \rf{sP5q} describes 
a system which is neither Painlev\'e III nor V equation  
but a new equation with only $A^{(1)}_1$ symmetry. 
It will be shown that equation \rf{sP5q} passes the Painlev\'e test
for all values of its parameters. Thus our study 
indicates that Dirac reduction of Hamiltonian systems
preserves the Painlev\'e property.
By letting one of these parameters $\epsilon_0$ or $r_1$ 
go to zero one recovers
Painlev\'e V or Painlev\'e  III equations. 
More specifically, in the limit $\epsilon_0 \to 0$ the  equation 
\rf{sP5q} becomes the Painlev\'e V equation with  $A^{(1)}_3$ 
extended affine Weyl symmetry  after transformation
from $q$ to $y =1 -r_1/q$.
On the other hand, in the limit $r_1 \to 0$ the  equation \rf{sP5q} goes into 
the Painlev\'e III equation invariant under 
$B^{(1)}_2$ extended affine Weyl symmetry. 



The schematic diagram of our presentation is shown below:
\vspace{5mm}

\includegraphics{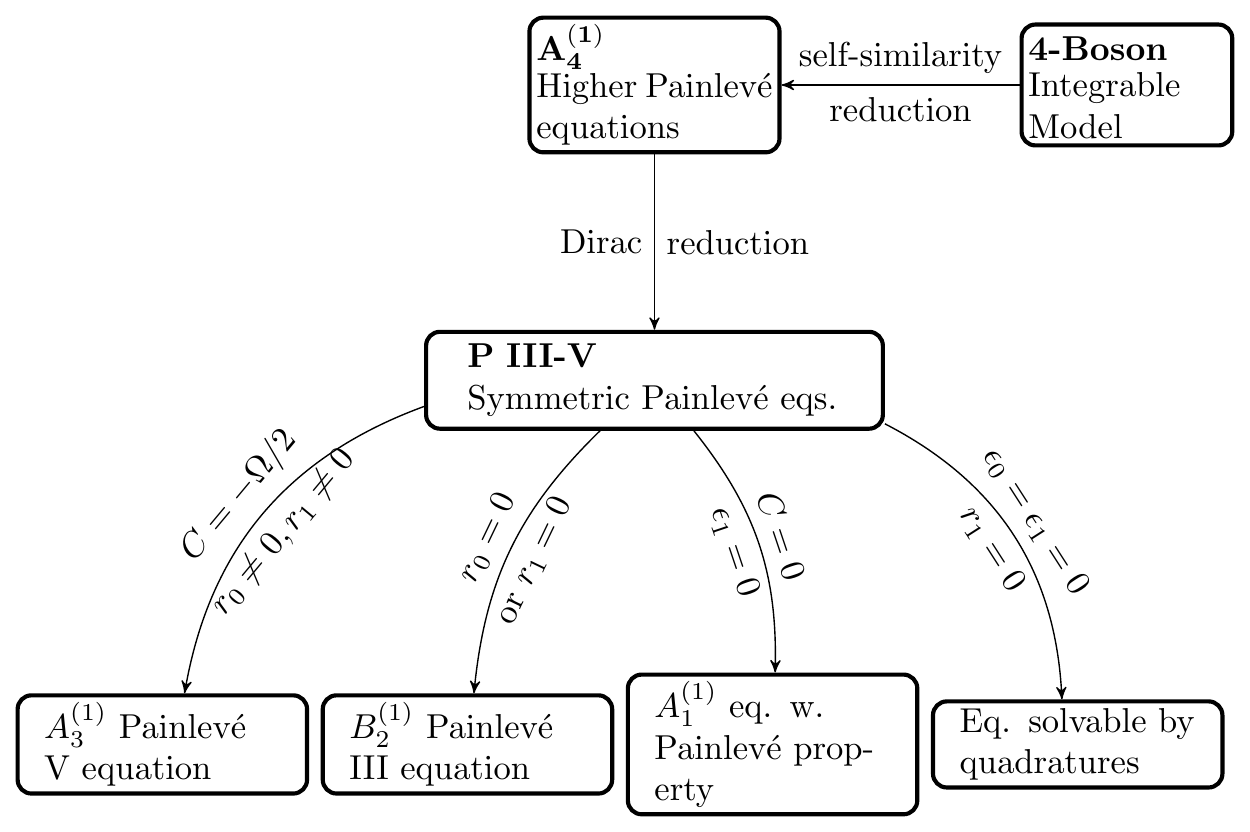}
\vspace{5mm}

{The above diagram  summarizes main results of 
our construction. 
Each of the Painlev\'e equations listed at the bottom of the 
diagram and associated with $A^{(1)}_3$, $B^{(1)}_2$, $A^{(1)}_1$
symmetry structures can be identified with reductions of 
higher Painlev\'e equations of $A^{(1)}_4$ type
via the Dirac Lagrange multiplier method within the Hamiltonian framework.
All these equations  can be embedded in a new set of symmetric 
symmetric Painlev\'e III-V equations and then recovered by taking
appropriate limits of the underlying parameters 
$C,r_i,\epsilon_i, i=1,2$.}

In section \ref{section:self} we obtain  higher  Painlev\'e
equations in a self-similarity limit of the $4$-boson integrable model
and recall their invariance under the extended affine Weyl 
$A^{(1)}_4$ group of B\"acklund transformations.

In section \ref{section:Dirac} we work within the Hamiltonian formalism 
applying Dirac Lagrange multiplier method \cite{claudio} on $A^{(1)}_4$ model
to obtain {Hamiltonian} structures governing 
symmetric Painlev\'e III-V
model. 

{
In section \ref{section:symmetric} we discuss symmetric Painlev\'e III-V equations 
and their reductions. In subsection \ref{subsection:a13painleve} 
we introduce a version 
of symmetric Painlev\'e V equations with explicit
dependence on an arbitrary constant $C$ that determines
form of the B\"acklund automorphism $\pi$ of the  extended affine
$A^{(1)}_{3}$ Weyl group.
By adding additional  terms to symmetric Painlev\'e V equations
that break ambiguity in the value of $C$ we arrive at the notion of 
symmetric Painlev\'e III-V equations presented  in 
subsection \ref{subsection:p3p5eqs}.
Various submodels obtained by setting  the underlying
constants to specific values and corresponding to 
symmetry structures, $A^{(1)}_3$, $B^{(1)}_2$, $A^{(1)}_1$, 
as well as  a solvable model,
are discussed in separate subsections of this
section. 
First, in \ref{subsection:p3p5eqs} we point out conditions,
$C=-\Omega/2, r_0 \ne 0, r_1 \ne 0$, on $C,r_1,r_2$
under which symmetric Painlev\'e III-V equations reduce to 
Painlev\'e V equations. In subsection \ref{subsection:B12}
we reduce symmetric Painlev\'e III-V equations to symmetric 
 Painlev\'e III equations in the $r_1 \to 0$ limit for any $C$ and
study emergence of $B^{(1)}_2$ symmetry group out of $A^{(1)}_3$ in this 
limit.
Hamiltonian representations of symmetric Painlev\'e III-V equations 
are introduced in subsection \ref{subsection:Hamiltonian}. In 
\ref{subsubsection:c=momega2}  it shown how for $C=-\Omega/2$ 
all Hamiltonian equations can be derived from one model
with the fixed value of $C$ taken here to be equal to $-1/2$. 
Thus, the Dirac reduction of Section \ref{section:Dirac} carried out
for $C=-1/2$ (and $C=0$) can be 
considered established for all values of $C=-\Omega/2$.
In \ref{subsubssection:notomega} and \ref{subsubsection:czero} 
we extend the Hamiltonian representations to
cases of $C\ne -\Omega/2$. Especially  in \ref{subsubsection:czero}
we cover the case of $C=0$ and  perform the Painlev\'e analysis of  
$A^{(1)}_1$ equation \rf{sP5q} obtained for  $C=0, r_0=1$ and $\epsilon_1=0$
that is a mixture of Painlev\'e III and 
V equations.
}

In a final section \ref{section:dicussion} we offer concluding remarks and 
discuss planned extensions of this work.

\section{Self-similarity reduction of $\mathbf{4}$-Boson integrable model to 
$\mathbf{A^{(1)}_4}$ Painlev\'e system}
\label{section:self}

The integrable $4$-boson model is defined by the pseudo-differential Lax
operator :
\begin{equation}
\begin{split}
L&= 
\left(\frac{\partial}{\partial x}-e_2 \right)
\left(\frac{\partial}{\partial x}-e_1-c_2\right)
\left(\frac{\partial}{\partial x}-c_1-c_2\right)  \\ &\times \left(\frac{\partial}{\partial x}-e_1-c_1-c_2\right)^{-1}
\left(
\frac{\partial}{\partial x}-e_2-c_2\right)^{-1},
\end{split}
\lab{4blax}
\end{equation}
and is a  particular case of the $2m$-boson sub-hierarchy 
of the KP hierarchy  with Lax operators that are given by a ratio of products :
\[
L_{m} = {\prod_{j=1}^{m+1} 
\left( \frac{\partial}{\partial x} + v_{m+2-j} \right)} 
\prod_{l=1}^{m}
\left( \frac{\partial}{\partial x} + \tilde{v}_l\right)^{-1},\quad
m=1,2, ...\, .
\]
The Lax coefficients of this sub-hierarchy are subject to the constraint :
\[
\sum_{j=1}^{m+1} v_j - \sum_{l=1}^{m} \tilde{v}_l= 0
\]
and the underlying second bracket structure 
has a form of graded $SL (m+1,m)$ Kac-Moody algebra in a diagonal gauge.
As shown in \cite{ANPZ,AGZ:2011}
the second bracket structure is diagonalized 
by a change of variables to $m$ conjugated pairs $\left( c_i \, ,\,
e_i\right)_{i=1,{\ldots} ,m}$ through relations
\[ 
\tilde{v}_l = - e_l - \sum_{p=l}^{m} c_p,     \quad 
v_j = - e_{j-1} - \sum_{p=j}^{m} c_p
\]
for $l =1,2, \ldots, m$ and $j=1, \ldots, m+1$.
These relations automatically solve the constraint and re-produce 
the graded $SL (m+1,m)$ bracket structure  
from a simple bracket structure
\begin{equation}
\left\{ e_i (x, [t] ), c_j (y, [t] )\right\}_2 = - \d_{ij} \d_x (x-y), \;\;
i,j =1,2, {\ldots}, m
\lab{2ndKP}
\end{equation}
where coefficients are taken at equal higher times $[t] =t_2, t_3,{\ldots}$
 \cite{ANPZ,AGZ:2011}. 

The $2m$-boson sub-hierarchy is characterized by invariance under 
Darboux-B\"acklund symmetry transformations associated with a special Volterra lattice 
symmetry structure within the bigger structure of Toda lattice
\cite{ANPZ}.


The 
system of partial differential equations  describing the second $t_2$- flow of the $4$-boson integrable hierarchy \cite{ANPZ}:
\begin{equation}
\lab{t2flow}
\begin{split}
\frac{\partial c_1}{\partial t_2}
&=\frac{\partial}{\partial x}\biggl(c_{1,x}-c_1^2-2e_1c_1+2c_{2,x}-
2c_1c_2\biggr)  \\
\frac{\partial e_1}{\partial t_2}&=\frac{\partial}{\partial x}
\biggl(-e_{1,x}-e_1^2-2e_1c_1-2e_1c_2\biggr) \\
\frac{\partial c_2}{\partial t_2}&=\frac{\partial}{\partial x}
\biggl(c_{2,x}-c_2^2-2e_2c_2\biggr) \\
\frac{\partial e_2}{\partial t_2}&=\frac{\partial}{\partial x}
\biggl(-e_{2,x}-e_2^2-2e_2c_2-2(e_{1,x}+c_1e_1)\biggr) 
\end{split}
\end{equation}
reduces in a self-similarity limit to ordinary differential 
equations with a
Painlev\'e property, in agreement with the ARS conjecture \cite{ARS}. 
The Painlev\'e equations obtained from equations
\rf{t2flow} in such limit will be a subject of this section.

To perform self-similarity reduction we introduce a  variable
$\xi={x}/{ \sqrt{t_2}}$ and define $\tilde{e}_j(\xi )$ and $ \tilde{c}_j(\xi )$
through relations
\[
\tilde{c}_j(\xi )= c_j(x,t_2) {\sqrt{t_2}},   
\hspace{1.5 cm} \tilde{e}_j(\xi)= e_j(x,t_2) {\sqrt{t_2}},
\qquad j=1,2\, .
\]
Equations \rf{t2flow} in a self-similarity limit
simplify when rewritten in terms of
$Y_1=\tilde{e}_1+\tilde{c}_1+2\tilde{c}_2$,  
$Y_2=\tilde{e}_2+\tilde{c}_2$ and 
$\tilde{e}_i (\xi), i=1,2$,
that  in this limit form two pairs of conjugated 
canonical variables.
To further  streamline notation we 
will drop the tilde over $\tilde{e}_i (\xi)$. 
In this notation a self-similarity limit of equations \rf{t2flow} 
takes a form of Hamilton equations:
\begin{equation}
e_{j,\xi}=
\frac{\partial{\cal H}_{A_4^{(1)}}}{\partial Y_j} , 
\hspace{1 cm} Y_{j,\xi}=-
\frac{\partial{\cal H}_{A_4^{(1)}}}{\partial {e}_j},  \hspace{1 cm}  j=1,2\, .
\lab{Hameqs}
\end{equation}
obtained from the Hamiltonian
\begin{align}
{\cal H}_{A_4^{(1)}}\left(e_1,
Y_1,e_2,Y_2 \right) &=-\sum_{j=1}^2 {e}_j \left({Y}_j-
\frac{\xi}{2}\right) \left({Y}_j-{e}_j\right)+2{e}_1 \left({Y}_1-
\frac{ \xi }{2}\right) \left({Y}_2-{e}_2\right) \nonumber \\
&+
\bar{k}_1 Y_1+\bar{k}_2 Y_2 -k_1{e}_1-k_2{e}_2. \lab{HA14}
\end{align}
with arbitrary constants $k_i, {\bar k}_i, i=1,2$.
The Hamilton equations are consistent with the following 
bracket structure
\begin{equation}
\{ e_i, {Y}_j \}= \delta_{ij}, \qquad i,j=1,2\, .
\lab{eYbracket}
\end{equation}

The Hamiltonian \rf{HA14} can be rewritten as \cite{agz}:
\begin{equation} \lab{calhmpq}
{\cal H}_{A^{(1)_4}} = \sum_{j=1}^2 p_j q_j  \left( p_j+q_j + \xi/2\right)
+2  p_1 q_1 p_2
-\sum_{j=1}^2 {\alpha}_{2j} q_j + \sum_{j=1}^2 p_j \left(\sum_{k=1}^j
\alpha_{2k-1}\right)
\end{equation} 
by employing a symplectic transformation:
\begin{equation}
\begin{split}
e_2 &= p_2+q_2+\xi/2, \qquad
e_{1} = -q_{1}\\
Y_2 &=  p_2+q_{1}+\xi/2, \qquad
Y_{1} =-q_{2}-p_2-p_{1}
\end{split}
\lab{eMMkr}
\end{equation}
and setting :
\[ \alpha_1=-{\bar k}_1, \;\; \alpha_2=-k_1-{\bar k}_2,
\;\; \alpha_3= {\bar k}_2 -k_2, \;\;\alpha_4= k_2 +{\bar k}_1,
\]
We  now can identify a self-similarity limit of 
\rf{t2flow} with  $A^{(1)}_4$ Painlev\'e equations:
\begin{equation}\lab{A2M}
 f_{i , \xi} = f_i \left(f_{i+1}- f_{i+2} + f_{i+3} - f_{i+4} \right) + \alpha_i, \quad i=0,\ldots,4 
\end{equation}
realized from a polynomial Hamiltonian \rf{calhmpq} 
through relations
\begin{equation}
p_i = f_{2i}, \qquad q_i= \sum_{k=1}^i f_{2k-1},\qquad  i=1,{\ldots}
,4 \, ,
\lab{pqf}
\end{equation}
when conditions $f_0 +\cdots+f_4=-\xi/2$ and
$\alpha_0+\cdots+ \alpha_4=-1/2$ are satisfied.

The above  Painlev\'e equations \rf{A2M} are manifestly invariant under
the following B\"acklund transformations: 
\begin{xalignat}{2}
s_i(\alpha_i)=-\alpha_i, &\quad  s_i(\alpha_j)=\alpha_j+\alpha_i~(j=i 
\pm 1), & s_i(\alpha_j)=\alpha_j~(j \ne i,i \pm 1),\nonumber \\ 
s_i(f_i)=f_i, & \quad s_i(f_j)=f_j \pm \frac{\alpha_i}{f_i}~(j=i \pm 1), 
& s_i(f_j)=f_j ~(j \ne i,i \pm 1),\nonumber \\
 \pi(\alpha_j)=\alpha_{j+1}, &\quad  \pi(f_j)=f_{j+1}, &
 i,j=0,1,{\ldots} ,4
\lab{BT:f}
\end{xalignat}
of the extended affine Weyl group ${A^{(1)}_{4}}$. The
generators $\pi, 
s_i,i=0,1,{\ldots}, 4$  satisfy the following 
fundamental relations 
\cite{Adler,NoumiYamada98a,NoumiYamada98b,Noumiwkb,noumibk}:
\begin{xalignat}{2}
   s_i^2=1, &\qquad \quad s_i s_j =s_j s_i~(j \ne i,i \pm 1), &
   s_i s_j s_i =s_j s_i s_j~(j=i \pm 1),\nonumber \\
  \pi^4=1, & \qquad \quad \pi s_j =s_{j+1}\pi.&
\lab{fun.rel}
\end{xalignat}
%
\section{Dirac Lagrange Multiplier Reduction of 
$\mathbf{{\cal H}_{A_4^{(1)}}}$ }
\label{section:Dirac}

Our starting point is the Hamiltonian ${\cal H}_{A_4^{(1)}}$ \rf{HA14}
describing a self-similarity  reduction of the 4-Boson model.

We first define a pair of canonical variables $P,Q$ :
\begin{equation}
P=- {N \xi} e_1 , \qquad Q= \frac{ Y_1-M \xi}{ N \xi}
\lab{p1q1}
\end{equation}
via symplectic transformation from $e_1,Y_1$ involving constants
$M,N$ and  a  variable $\xi$ from previous section. 
This change of variables redefines the corresponding Hamiltonian
${\cal H}_{A_4^{(1)}}$ as follows:
\[ 
{\cal H}_{A_4^{(1)}} (P,Q,e_2,Y_2) = 
{\cal H}_{A_4^{(1)}} \left(e_1,
Y_1,e_2,Y_2 \right) 
-\frac{PQ}{\xi} - \frac{M P}{N \xi}
\]
so that ${\cal H}_{A_4^{(1)}}$  in terms of $e_2,Y_2, P, Q$ becomes:
\[
\begin{split}
&{\cal H}_{A_4^{(1)}} (P,Q,e_2,Y_2) =- \frac{1}{(2\,\xi\,N^{2}) }
( - 2\,M\,P^{2} + P^{2} - 2\,{{\bar k}_2}\,
{Y_2}\,N^{2}\,\xi - 2\,{k_1}\,P\,N\\
&+ 2\,{k_2}\,
{e_2}\,N^{2}\,\xi + 2\,P\,Q\,N^{2} 
 + 2\,M\,P\,N - 2\,P\,Q^{2}\,N^{3}\,\xi^{2} \\
 &- 2\,P\,M^{2}\,\xi^{2}\,N + P\,
Q\,N^{2}\,\xi^{2} + P\,M\,\xi^{2}\,N - 2\,Q\,N\,P^{2}  \\
& + {e_2}^{
2}\,N^{2}\,\xi^{2}- 4\,P\,Q\,N^{2}\,M\,\xi^{2} + 2\,{e_2}\,N^{2}\,\xi\,
{Y_2}^{2} - 2\,{e_2}^{2}\,N^{2}\,\xi\,{Y_2}\\ 
&- 
{e_2}\,N^{2}\,\xi^{2}\,{Y_2}
 - 2\,P\,\xi\,N\,{Y_2}+ 2\,P\,\xi\,N\,{e_2} - 2\,{{\bar k}_1}\,\xi^{2}\,N^{3}
\,Q \\
&+ 4\,P\,\xi\,N^{2}\,Q\,
{Y_2} - 4\,P\,\xi\,N^{2}\,Q
\,{e_2}+ 4\,P\,\xi\,N\,M\,{Y_2} - 4\,P\,\xi\,N\,M\,{e_2}
)
\end{split}
\]
up to constant terms.

To eliminate variables $e_2$ and $Y_2$ we impose two second class constraints :
\begin{subequations}\lab{phi}
\begin{align}
\phi_1 &= e_2 - \frac{1}{\xi} \left( D \xi^2 + E P +F P Q \right)
\lab{phi1}\\
\phi_2 &= Y_2 -A \xi \lab{phi2} 
\end{align}
\end{subequations}
with constants $D, E, F,A$ (together with $N,M$ introduced earlier) 
to be completely determined by 
the condition that the reduction process reproduces 
one of the following Hamiltonian systems :
\begin{itemize}
\item The Hamiltonian system that will be shown to describe 
equation \rf{big} for $C=0, r_0=1,{\Omega=1}$ and $\epsilon_1=0$
and $q=f_1,p=-f_2$ as in relation \rf{fpdict}:
\begin{equation}
z H_{C=0} = p (p+z)q(q-r_1) +\alpha_2 z q +(\alpha_1 r_1 
+\epsilon_0 z) p-
(\alpha_1+\alpha_3) pq
\lab{sasanohame}
\end{equation}
\item The Hamiltonian system that will be shown to describe equation \rf{big} 
for $C=-1/2,{\Omega=1}, r_0=1$ and $\epsilon_1=-1$
and for $q = f_1 z^{-1/2}, p = -f_2 z^{1/2}$ as in relation
\rf{fpzdict}:
\begin{equation}
\begin{split}
z H_{C=-1/2} &= p  \left(p+z \right) q  \left(q-r_1\right)-
\left(\alpha_1+\alpha_3\right) pq
+\left(\alpha_1 r_1 +\epsilon_0\right) p 
\\&+ \left( \alpha_2 -
r_1 \right) z q
\end{split}
\lab{hamChalf}
\end{equation}
\end{itemize}
Here we define $z$ as $z=\xi^2$.
Below in subsection \ref{subsection:Hamiltonian} 
we will discuss how these 
two Hamiltonian structures fit into the 
formalism of symmetric PIII-V equations {and show
that all the Hamilton equations  with $C=-\Omega/2$ can be obtained
from $H_{C=-1/2}$ by simple rescaling.}

We will now follow the Lagrange multiplier approach to impose the
constraints from \rf{phi} by augmenting the Hamiltonian
${\cal H}_{A_4^{(1)}}$ by Lagrange multiplier terms:
\[{\cal H}_{A_4^{(1)}} \to {\cal H}_{A_4^{(1)}}^\lambda  =  {\cal H}_{A_4^{(1)}}
+ \lambda_1 \phi_1 + \lambda_2 \phi_2\, .
\]
The Lagrange multipliers $\lambda_1, \lambda_2$ are fixed 
by condition of compatibility of constraints $\phi_1, i=1,2$:
\begin{equation}
0= \dder{\phi_m}{\xi} = \{ \phi_m, H_{A_4^{(1)}} \} 
+ \pder{\phi_m}{\xi}+ \sum_{n=1}^2 \lambda_n \{ \phi_m, \phi_n \}
\lab{compatibility}
\end{equation}
on the constraint manifold. These compatibility conditions
fix the values of Lagrange multipliers to:
\begin{equation}
\lambda_n= - 
\sum_{n=1}^2 \{ \phi_n, \phi_m \}^{-1} 
\left( \{ \phi_m, H_{A_4^{(1)}} \} 
+ \pder{\phi_m}{\xi} \right)
\lab{lambdan}
\end{equation}
where the values of the matrix elements of  the inverse of the matrix $\{ \phi_n, \phi_m \}$
are $\{ \phi_1, \phi_2 \}^{-1}=-1 = - \{ \phi_1, \phi_2 \}^{-1} $ and
zero otherwise.
Thus
\[ {\cal H}_{A_4^{(1)}}^\lambda  =  {\cal H}_{A_4^{(1)}} - \sum_{n=1,m=1}^2
 \phi_n \{ \phi_n, \phi_m \}^{-1} 
\left( \{ \phi_m, H_{A_4^{(1)}} \} 
+ \pder{\phi_m}{\xi} \right)
\]
and equations of motion for $P,Q$ become
\begin{subequations}\lab{varxi}
\begin{align}
P_\xi &= \{ P, H_{A_4^{(1)}} \} -  \{ P, \phi_1 \}
\{ \phi_1, \phi_2 \}^{-1}\left( \{ \phi_2, H_{A_4^{(1)}} \} 
+ \pder{\phi_2}{\xi} \right) \vert_{\phi_i \approx 0}
\lab{Pxi} \\
Q_\xi &= \{ Q, H_{A_4^{(1)}} \} -  \{ Q, \phi_1 \}
\{ \phi_1, \phi_2 \}^{-1}\left( \{ \phi_2, H_{A_4^{(1)}} \} 
+ \pder{\phi_2}{\xi} \right) \vert_{\phi_i \approx 0}
\lab{Qxi} 
\end{align}
\end{subequations}
Explicit calculation gives the following values for quantities appearing
in \rf{varxi}:
\begin{equation}
\{ \phi_2, H_{A_4^{(1)}} \} = - \pder{H_{A_4^{(1)}}}{e_2}, \quad
\{ P, \phi_1 \} = \frac1\xi F P, \quad
\{ Q, \phi_1 \} = - \frac1\xi \left( E +F Q\right) \, .
\lab{PQphi}
\end{equation}
Multiplying both sides of equations \rf{varxi} by $\xi$ in order
to obtain  expressions for $ \xi P_\xi, \xi Q_\xi$ and inserting 
the technical results \rf{PQphi} yields:
\begin{subequations}\lab{xivarxi}
\begin{align}
\xi P_\xi &= - \xi \pder{H_{A_4^{(1)}} }{Q} - F P \left( 
\pder{H_{A_4^{(1)}} }{e_2} +A   \right) \vert_{\phi_i \approx 0}
\lab{xiPxi} \\
\xi Q_\xi &= \xi \pder{H_{A_4^{(1)}} }{P}  +\left( E+FQ\right)
\left( \pder{H_{A_4^{(1)}} }{e_2} +A   \right) \vert_{\phi_i \approx 0}
\lab{xiQxi} 
\end{align}
\end{subequations}

We will first compare equations \rf{xivarxi}
to the Hamilton equations :
\begin{equation}
\begin{split}
z q_z &= q(q-r_1) (2p+z)- (\alpha_1+\alpha_3) q+ \alpha_1 r_1+
\epsilon_0 z\\
zp_z &= -  p (p+z)(2q-r_1) + (\alpha_1+\alpha_3) p- \alpha_2 z
\end{split}
\lab{eqsmots}
\end{equation}
obtained from $H_{C=0}$ given in
\rf{sasanohame}. For $z=\xi^2$ these equations can be rewritten as
\begin{equation}
\xi q_{\xi}= 2 \pder{H_{C=0}}{p},
\quad \xi p_{\xi}= - 2 \pder{H_{C=0}}{q}\, .
\lab{heqsmots}
\end{equation}
Equations  \rf{xiPxi},\rf{xiQxi} will agree
with  equation \rf{heqsmots} when we identify $P=p$ and $Q=q$ 
and values
 of $N,M, D, E, F,A$ become functions of $r_1$ and $\epsilon_0$
that will now be  given below.
We find that 
\[ N=2, 
\qquad A= \frac12 +\frac{32 \epsilon_0}{16 r_1^2-1}
\]
and $F \ne 0$ is a  solution of a quadratic equation:
\begin{equation}
16 \epsilon_0 F^2+(16 r_1^2-1)( F-1)=0 \, .
\lab{Feq}
\end{equation}
Note, that for $16r_1^2-1=0$ or $r_1= \pm 1/4$ it must hold that 
$\epsilon_0=0$.
It is convenient to express the remaining parameters $M, D, E$ in terms of
$F$ being a solution \rf{Feq} and an auxiliary quantity:
\[ 
G= 1-F+4 r_1^2 F^2 = F^2 \frac{\epsilon_0^2+r_1^2/4-4r_1^4}{1/16-r_1^2}
\]
as 
\begin{align}
D&= \frac{1}{F^2 }\left(-F+F^2/4+1 \pm \frac12 \sqrt{G}\right)\\
&=\frac12 \left(\frac12 +\frac{32 \epsilon_0}{16r_1^2-1}\right)
\pm \frac12 \sqrt{\frac{\epsilon_0^2+r_1^2/4-4r_1^4}{1/16-r_1^2}} \\
M&=\frac{1}{F^3} \left( (1-2r_1) F^3-2F^2+2F \pm F(F-2) \sqrt{G}\right)\\
E&=-\frac{1}{4 F} \left(2r_1 F^2+F \pm F  \sqrt{G}\right)
\end{align}
In addition the following conditions on constants $k_1,k_2, {\bar k}_1
$ need to hold :
\begin{align}
{\bar k}_1 &= \alpha_2 \\
k_2&= \frac{1}{2F^2} \left(F^2+4-6F_+4 F(\alpha_1+\alpha_3)\right)\\
k_1 &=\frac{1}{2F^3} \left(F^3(1+4r_1(\alpha_1-\alpha_3)) +2F
-F^2(2(\alpha_1+\alpha_3)+1) +\right.\\
 & \left. \pm 2 F (F(\alpha_1+\alpha_3)-1)\sqrt{G} \right)\, .
\end{align}

Next we determine conditions 
for which equations  \rf{xivarxi} and the Hamilton equations 
\begin{equation}
\begin{split}
z q_z &= q \left(q-r_1\right) \left(2p+z\right)- 
\left(\alpha_1+\alpha_3\right) q
+\alpha_1 r_1+\epsilon_0\\
z p_z &=  - p \left(p+ z \right)\left(2q-r_1\right) + 
(\alpha_1+\alpha_3) p - \alpha_2  z + r_1 z
\end{split}
\lab{Hamceqs}
\end{equation}
will agree.   Equations \rf{Hamceqs} are 
 obtained from the Hamilton $H_{C=-1/2}$ (see definition 
 \rf{hamChalf}), which as will {be} shown in the next section describes 
\rf{big} for a special non-zero value of the parameter $C$ ($C=-1/2$)
and $r_0=1,  \epsilon_1=-1$.
Since we identify $z=\xi^2$ the comparison is  to be made with equations
\begin{equation}
\xi q_{\xi}= 2 \pder{H_{C=-1/2}}{p},
\quad \xi p_{\xi}= - 2 \pder{H_{C=-1/2}}{q},
\lab{hHamceqs}
\end{equation}
This procedure fixes the values of $N,M, D, E, F,A$ to
 \[F=1,\quad A=\frac12,\quad N=2,\quad D=\frac14\pm r_1
 \]
 and
 \[
 E = \left\{ \begin{matrix}- \frac{1}{4}\\
 -(r_1+1/4) \end{matrix} \right. , \; \qquad
 M= \left\{ \begin{matrix} 1/2-2  r_1 \\
 1/2 \end{matrix} \right.
 \]
 and
 \begin{align}
 {\bar k}_1 &= \alpha_2 - r_1\\
 k_2&= \frac{3}{2} - 2\left(\alpha_1+\alpha_3)\right)\\
 k_{1 \, \pm} &= \alpha_0 +\alpha_2+2 r_1(2 \alpha_1-1)+
 4 \epsilon_0
 +
 \left\{ \begin{matrix} 0 \\
 4 (\alpha_0+\alpha_2)r_1  \end{matrix} \right.
 \end{align}


In conclusion, we have obtained  two 
Hamilton
equations {\rf{eqsmots} and \rf{Hamceqs}} by Dirac reduction
of the self-similarity limit of the $4$-boson model. 
In the next section
the Hamilton equations {\rf{eqsmots} and \rf{Hamceqs}} 
obtained in this section will be identified
with generic examples of new symmetric PIII-PV equations.

\section{Symmetric Formulation of Painlev\'e III and V systems}
\label{section:symmetric}
\subsection{On Symmetric $\mathbf{A^{(1)}_{3}}$ Painlev\'e equations}
\label{subsection:a13painleve}
To provide a useful framework for combining reductions of systems described
in the previous section  we first recall 
symmetric  $A^{(1)}_{3}$  Painlev\'e V (PV) equations (see e.g. 
\cite{NoumiYamada98a}) :
\begin{align}
 z  f_{i, \,z} &=f_i f_{i+2} \big( f_{i+1}-f_{i+3}
\big)+(-1)^{i} f_i \,\big(\alpha_{1}+\alpha_{3}+C\big) \lab{A2M+1a}
\\ 
&+
\alpha_i\big(f_{i}+f_{i+2}\big),\qquad \qquad \qquad i=0,1,2,3\, ,
\nonumber
\end{align}
where $C$  and $\alpha_i, i=0,1,2,3$
are being constants. Unlike the reference \cite{NoumiYamada98a}
we do not assume
relation \rf{Calphas} between  $C$  and $\Omega$.

Equations \rf{A2M+1a} are manifestly invariant under 
the extended affine Weyl group ${A^{(1)}_{3}}$ generators $\pi, 
s_i,i=0,1,2,3$ acting as defined in relations \rf{BT:f}.
However the action of the automorphism operator $\pi$  depends on whether 
a parameter $C$ satisfies the condition \rf{Calphas}
as we will now discuss.

{}By summing  $i=0,2$  and $i=1,3$ components of 
equation  \rf{A2M+1a} one obtains 
\begin{equation}
\begin{split}
z  \dder{(f_0+f_2)}{z} &=
(f_0+f_2)(C+\alpha_0+\alpha_1+\alpha_2+\alpha_3),
\\ z  \dder{(f_1+f_3)}{z} &= (f_1+f_3) (-C) \, ,
\end{split}
\lab{condc}
\end{equation}
which can be rewritten as equations \rf{condcbig}, 
revealing two integration constants $r_0$ and $r_1$ of equation
\rf{A2M+1a}.
Note that having identical equation $z (f_i+f_{i+2})_z =
(-C) (f_i+f_{i+2} )$ 
for both $ i=0 $ and $i=1$
is a necessary condition for canonical ($\pi(f_i)=f_{i+1}$) transformation rule
for $\pi$. This requires  the condition \rf{Calphas} with $r_0=r_1$.
When the condition \rf{Calphas} 
holds one can cast equations \rf{condcbig}
in a compact form ;
\begin{equation}
f_i+f_{i+2}= r_i z^{-C}, \qquad \; i=0,1 \, .
\lab{condcbig1}
\end{equation}
 
We also observe that the  substitution :
\begin{equation} {\bar f}_j = z^{\kappa} f_j, \;\;j=1,3  
\qquad {\bar f}_i =  f_i/ z^{\kappa},
\;\; i=0,2 \lab{subskappa}
\end{equation}
in equation \rf{A2M+1a} 
shifts the value of $C$ to
${\bar C}= C-\kappa $ and changes the canonical  action of $\pi$ automorphism to:
\[  \pi({\bar f}_{2j})=  \frac{1}{z^{2 \kappa}} {\bar f}_{2j+1}, \;\;
\pi({\bar f}_{2j-1})= z^{2 \kappa} {\bar f}_{2j}
\]
Despite its non-conventional
form such operator $\pi$ continues to satisfy the extended affine 
Weyl group fundamental relations \rf{fun.rel}.
In particular, for the choice $\kappa=C$ the value of
${\bar C}= C-\kappa$ becomes equal to zero.
One should add that setting  explicitly $C=0$ in equation
\rf{A2M+1a} with $\Omega=1, r_0=1$ as it was done in \cite{sasano} 
results in invariance  under  
$\pi$ given by
\begin{equation}
 \pi (\alpha_i) = \alpha_{i+1}, \;\; \pi (f_{2i}) = \frac{z}{r_1} 
 f_{2i+1},\;\;
 \pi (f_{2i-1}) = \frac{r_1}{z} f_{2i}\, ,
 \lab{sasanopi}
 \end{equation}
which agrees with above discussion on consequences 
of shifting the
value of $C$ in \rf{A2M+1a} for the form of automorphism $\pi$.

Thus for the symmetric {Painlev\'e} V equation \rf{A2M+1a} one is able to vary the value of the parameter $C$. The equation \rf{A2M+1a} 
remain invariant under the extended affine Weyl symmetry although
such transformations of $C$ modify
a form of the  automorphism $\pi$. 

\subsection{Symmetric PIII-PV equations}
\label{subsection:p3p5eqs}
In this subsection, we propose a model that combines  Painlev\'e III
and  Painlev\'e V equations and study its properties. 

We start our discussion by noticing
an obvious ambiguity in the definition \rf{A2M+1a}.
For a fixed value of  
$\alpha_0+\alpha_1+\alpha_2+\alpha_3= \Omega \ne 0\, ,$
let us consider adding additional terms  
$(-1)^{[i/2]} \kappa_i \big(f_{i}+f_{i+2}\big)$
in \rf{A2M+1a} :
\begin{align}
z f_{i,\, z} &=f_i f_{i+2} \big(
 f_{i+1}- f_{i+3} \big)+
(-1)^{i} f_i \,\big(\alpha_{1}+\alpha_{3}+C\big)  \lab{ambiguity}
\\ 
& +
\alpha_i\big(f_{i}+f_{i+2}\big)
-(-1)^{[i/2]} \kappa_i \big(f_{i}+f_{i+2}\big),
\nonumber
\end{align}
where 
$\kappa_i=\kappa_{i+2}$. Despite the presence of additional terms 
with $\kappa_i$ equation  \rf{ambiguity} remains invariant under
the extended affine Weyl group ${A^{(1)}_{3}}$ with transformations 
\rf{BT:f} as long as one substitutes 
$\alpha_i$ by $\bar{ \alpha_i} =\alpha_i -(-1)^{[i/2]} \kappa_i$.

We have seen above that the parameter $C$ in \rf{A2M+1a} can be shifted 
by a simple re-scaling of $f_i$'s that keeps \rf{A2M+1a} invariant but 
in the process changes the form of action of $\pi$.

In order to fix the value of the parameter $C$
we now define a system shown in \rf{big}, which for the convenience
of the reader we reproduce here:
\begin{align}
z f_{i,\,z} &=f_i  f_{i+2} \big(
 f_{i+1}- f_{i+3} 
 \big)+(-1)^{i} f_i \,\big(\alpha_{1}+\alpha_{3}+C\big)  \tag{1.1}
\\ 
&+
\alpha_i\big(f_{i}+f_{i+2}\big)
-(-1)^{[i/2]} \epsilon_{i+1} \big(f_{i+1}+f_{i+3}\big), \qquad
i=0,1,2,3\, .
\nonumber
\end{align}
There are two deformation parameters involved in this
construction; $\epsilon_0(=\epsilon_2)$ and 
$\epsilon_1(=\epsilon_3)$. 

The system \rf{big} is invariant under $\pi(\alpha_j)=\alpha_{j+1}, 
\pi(f_j)=f_{j+1}, \pi(\epsilon_j)=(-1)^{j+1} \epsilon_{j+1}$
but the equation \rf{big} is no longer invariant under
substitution \rf{subskappa} and therefore the value of the constant
$C$ can not be shifted arbitrarily through rescaling of $f_i$'s. 
The equations \rf{condc} still hold for the system \rf{big}
with integration constants $r_0,r_1$ defined as in \rf{condcbig}.

As long as the condition \rf{Calphas} ($C=-\Omega/2$)
holds and $r_1 \ne 0, r_0 \ne0$  
we can cast the system of equations \rf{big} into
\rf{ambiguity}
with $\kappa_1 = \epsilon_1  r_0/r_1$ and 
$\kappa_0 = \epsilon_0  r_1/r_0$. Thus for that case
the system  \rf{big} is equivalent to 
the symmetric Painlev\'e ${A^{(1)}_{3}}$ equation \rf{ambiguity}
with shifted parameters $\alpha_i \to {\bar \alpha}_i$:
\begin{equation}
{\bar \alpha}_i = \alpha_i -(-1)^{[i/2]} 
\frac{\epsilon_{i+1}r_{i+1}}{r_i}
\lab{shiftalpha}
\end{equation}
in B\"acklund relations \rf{BT:f}.

Although equation \rf{shiftalpha} is only valid for $r_i \ne 0$  
it also signals what to expect when one of the integration constants
goes to zero. If, for instance, $r_1 \to 0$ then the formula 
\rf{shiftalpha} diverges for ${\bar
\alpha}_1=\alpha_1+\epsilon_0r_0/r_1$ and 
${\bar \alpha}_3 =\alpha_3 -\epsilon_0r_0/r_1$
indicating breaking of $s_1, s_3$ symmetries.
Interestingly, the limit $r_1 \to 0$ can however be taken of $s_1$ and
$s_3$  acting successively in the 
product $s_1 s_3$. Observe namely that 
\begin{equation}
\begin{split}
s_1s_3 (f_2) &= f_2- \frac{{\bar \alpha}_3}{f_3}+
\frac{{\bar \alpha}_1}{f_1}=
f_2- \frac{\alpha_3+\epsilon_0r_0/r_1}{-f_1 + r_1 z^{-C}}+\\
&+ \frac{\alpha_1-\epsilon_0r_0/r_1}{f_1}\underset{r_1 \to 0}{\longrightarrow} 
f_2 +\frac{\alpha_1+\alpha_3}{f_1}+\frac{\epsilon_0r_0z^{-C}}{f_1^2}\\
s_1s_3 (f_1) &= f_1 , \quad s_1s_3 (\alpha_0) =
\alpha_0+\alpha_1+\alpha_3, \quad s_1s_3 (\alpha_2) =
\alpha_2+\alpha_1+\alpha_3
\lab{s1s3}
\end{split}
\end{equation}
Given these expressions we are able to verify that 
\begin{equation}
\lim_{r_1 \to 0} s_1 s_3 =\pi^2 \pi_0\pi_2 \lab{s12pi02}
\end{equation}
with explicit expressions for symmetry generators 
$\pi^2 ,\pi_0,\pi_2$ 
{that emerge below  in 
subsection \ref{subsection:B12} as part of the 
$B^{(1)}_2 $ symmetry structure obtained 
in the limit $r_1 \to 0$ for an arbitrary $C$.}

\subsection{ $\mathbf{B^{(1)}_2}$-Model for 
$\mathbf{r_1}$
being zero.}
\label{subsection:B12}
In this case we set one of the integration constants, chosen here
to be $r_1$, to zero.  Note, that setting $r_1=0$ effectively
leaves only one deformation parameter $\epsilon_0$ as
$\epsilon_1$ appears in equation \rf{big} only in the product with $r_1$.
For $r_1=0$ it follows  that $f_1+f_3=0$ and
$\alpha_3=\alpha_1=(\Omega-\alpha_0-\alpha_2)/2$.
Consequently, equation \rf{big} reduces to
\begin{subequations}\lab{symP3C}
\begin{align}
z \dder{f_0}{z}&=2 f_0 f_1f_2+ (2 \alpha_1 +C)f_0+\alpha_0 (f_0+f_2)
\lab{symP3Ca}\\
z \dder{f_1}{z}&=  f_1^2( f_0 -f_2)-
(2\alpha_1+C) f_1+\epsilon_0  (f_0+f_2) 
\lab{symP3Cb} \\
z \dder{f_2}{z}&= -2 f_2 f_0 f_1+(2\alpha_1 +C)f_2+\alpha_2 (f_0+f_2)
\, .
\lab{symP3Cc} 
\end{align}
\end{subequations}
The system \rf{symP3C} provides a symmetric representation of Painlev\'e 
III for an arbitrary parameter $C$. 

Note, that a similar structure appeared earlier in 
\cite{sasano0704} and this structure can be obtained from \rf{symP3C} by setting
$\eta=\epsilon_0$ and  introducing ${\bar f}_1 = \sqrt{t} f_1, \quad {\bar f}_i =  f_i/  \sqrt{t},
\;\; i=0,2$ for $C=-1/2, \Omega=1$.

We will now study symmetries of equations 
\rf{symP3C}. In addition to an obvious identity automorphism:
$f_i \to f_i, \alpha_i \to \alpha_i$ 
there also exists  a sign reversal automorphism:
$f_i \to - f_i, \alpha_i \to \alpha_i$ of order $2$.
A sign reversal automorphism requires a corresponding 
transformation of $z$ to be consistent with 
the second of conditions listed in
\rf{condcbig}. For example, for $C+\Omega=1$, the consistency requires 
$z \to -z$ to agree with the transformation $f_0,f_2 \to -f_0 , -f_2$.

Equations \rf{symP3C} are invariant under $\pi^2$ being
``the square root" of 
an identity automorphism defined on extended parameter space that includes 
$\epsilon_0$ :
\[\pi^2 :\;   f_1 \to - f_1, \; f_2 \to
f_0,\; f_0 \to f_2, \;\alpha_0 \to \alpha_{2}, \; 
\alpha_2 \to \alpha_0,\;
\epsilon_0 \to - \epsilon_0 \, .
\]
In addition,  equations \rf{symP3C} are invariant under ``the square 
root" of  a sign reversal automorphism  that involves transformations of $z$ with imaginary
parameters  :
\[ 
\rho :\;  f_1 \to i f_1 , \;
f_2 \to i f_{0}, \; f_0 \to i f_2, \;
\alpha_0 \to \alpha_{2}, \; 
\alpha_2 \to \alpha_0,\; z^{C+\Omega} \to i z^{C+\Omega} \,,
\]
and $\rho(\epsilon_0)=\epsilon_0$.
Note that $\rho$ is an automorphism of order $4$ and $\pi^2$ and $\rho$
commute
\[ \rho \; \pi^2 = \pi^2 \rho\,.\]

Equation \rf{symP3C} remains invariant under 
$s_0$ and $s_2 $ :
\begin{align}
s_0 &: f_1 \to  f_1+\frac{\alpha_0}{f_0}, \;
f_0 \to f_0 ,\; f_2 \to f_2 , \alpha_0\to -\alpha_0 ,\;
\alpha_2\to\alpha_2 ,\; \alpha_1 \to \alpha_1+\alpha_0 \\
 s_2 &: f_1\to f_1- \frac{\alpha_2}{f_2},\;
 f_0\to f_0\; f_2\to f_2 ,\;
 \alpha_0\to \alpha_0 ,\;
 \alpha_2\to -\alpha_2,\; \alpha_1 \to \alpha_1+\alpha_2 \, .
\lab{ss02}
\end{align}

For $\epsilon_0 \ne 0$ the system \rf{symP3C} is no longer invariant 
under $s_1,s_3$ transformations defined in equations \rf{BT:f},
{which together with $s_0,s_2$ were part of $A^{(1)}_3$ symmetry structure
of PV equation \rf{A2M+1a}. }
However for $\epsilon_0 \ne 0$
there emerge two  additional symmetries given by : 
\begin{align}
\pi_0 &: f_1 \to  -\frac{\epsilon_0}{f_1},\;
f_0 \to \frac{-1}{\epsilon_0 } \left( f_1^2 f_2 - \alpha_2 f_1 \right)
+ ( f_0+f_2) ,\;
f_2 \to \frac{f_1}{\epsilon_0 } \left( f_1 f_2 -\alpha_2\right)
,\nonumber \\
&\alpha_0 \to 2 \Omega+2C-\alpha_0 ,\;
\alpha_2\to\alpha_2  \lab{pi0} \\
 \pi_2 &: f_1\to  \frac{\epsilon_0}{f_1},\;
 f_0 \to -\frac{f_1}{\epsilon_0 } \left( f_1 f_0+ \alpha_0  \right) ,
 \;
 f_2 \to \frac{1}{\epsilon_0} \left(f_1^2 f_0 + \alpha_0 f_1\right)+
  (f_0+f_2 ) ,\nonumber \\
& \alpha_0\to \alpha_0 ,\;
 \alpha_2\to 2 \Omega +2C- \alpha_2 \, ,
\lab{pi2}
\end{align}
which  for $C=-\Omega/2$ are related to $s_1,s_3$ transformations 
of $A^{(1)}_3$-model
through a limiting procedure introduced in \rf{s12pi02}.


The B\"acklund transformations $s_0,s_2, \pi^2, \pi_0, 
\pi_2$ all square to one:
\[
s_0^2=1, \; s_2^2=1,\;  (\pi^2)^2=1, \;  \pi_0^2=1,\; \pi_2^2=1\, .
\]
while $\rho^4=1$. In addition they satisfy the following fundamental
B\"acklund relations:
\[
\pi^2 \left( s_0, s_2, \pi_0, \pi_2\right) =
\left( s_2, s_0, \pi_2, \pi_0\right) \pi^2 ,\;\;
\rho \left( s_0, s_2, \pi_0, \pi_2\right) =
\left( s_2, s_0, \pi_2, \pi_0\right) \rho
\]
as well as 
\begin{xalignat}{2}
\left( s_0 s_2 \right)^2 &=1,\; &\left( \pi_0 \pi_2 \right)^2 =1\,
\lab{Back1}\\
\left( \pi_0 s_2 \right)^2 &=1,\; &\left( \pi_2 s_0 \right)^2 =1\, .
\lab{Back2}
\end{xalignat}
Relations \rf{Back1} amount to commutativity
of $s_0$ with $s_2$ and $\pi_0$ with $\pi_2$:
\[
s_0 s_2= s_2 s_0\, , \;\; \pi_0 \pi_2= \pi_2 \pi_0\,,
\]
while relations \rf{Back2} are equivalent to
 commutativity
of $s_0$ with $\pi_2$ and $s_2$ with $\pi_0$ :
\[
s_0 \pi_2= \pi_2 s_0\, , \;\; s_2 \pi_0= \pi_0 s_2\,.
\]
The last two relations transform into each other under $\pi^2$
conjugation.
The reference \cite{sasano0704} introduced in a setting of symmetric
PIII equation 
a generator $s_1^s $, which in
our case corresponds to $\pi^2 \pi_0\pi_2$ for $C=-\Omega/2=-1/2$. 
The identities 
$(s_0s_1^s)^4=1$ and $(s_0s_1^s)^4=1$ in reference 
\cite{sasano0704} can easily be verified as direct 
consequences of \rf{Back1} and \rf{Back2}.


Below, we will discuss relation of the above symmetries to the extended affine Weyl
$B^{(1)}_2$ group. Let us inspect action of 
transformations $ s_2,\pi_0, \pi^2$: 
\begin{xalignat}{1}
\binom{\alpha_0}{\alpha_2} &\stackrel{s_2}{\longrightarrow}
\binom{\alpha_0}{-\alpha_2} \quad\quad\qquad\quad\; \to \quad \binom{v_1}{v_2} \stackrel{s_2}{\longrightarrow}
 \binom{v_2}{v_1}
\lab{s2vv}\\
\binom{\alpha_0}{\alpha_2} &\stackrel{\pi^2}{\longrightarrow}
\binom{\alpha_2}{\alpha_0}\;\;\quad \qquad \quad\quad\;\; \to \quad \binom{v_1}{v_2} \stackrel{\pi^2}{\longrightarrow}
 \binom{v_1}{-v_2} 
\lab{pivv}\\
\binom{\alpha_0}{\alpha_2} &\stackrel{\pi_0}{\longrightarrow}
\binom{2 \Omega  + 2C-\alpha_0}{\alpha_2} \;\quad\to \;  \binom{v_1}{v_2} 
\stackrel{\pi_0}{\longrightarrow}
\binom{-1-v_2}{-1-v_1}
\lab{pi0vv}
\end{xalignat}
on $v_1,v_2$ 
variables:
\begin{equation}
{v_1=    \frac{-1}{2(\Omega +C)} \left(\alpha_0+\alpha_2\right), \qquad
v_2= \frac{1}{2(\Omega +C)}  \left( \alpha_2-\alpha_0 \right) \,,}
\lab{witte-relsp}
\end{equation}
valid for $C\ne -\Omega$ (for $C=-\Omega$ the symmetry group simplifies
into action of two $A^{(1)}_1$ groups).

One sees that actions of $\pi_0, s_2,\pi^2$ on parameters
$(v_1,v_2)$  realize a representation of the extended affine Weyl group for the root
system $B^{(1)}_2$ \cite{okap3,witte,kmnoy,forrester}.
To see the connection to the $B^{(1)}_2$ root lattice 
consider a 2-dimensional vector space ${\mathbf V}$ consisting of vectors
${\mathbf v} = v_1 {\mathbf e_1} +v_2 {\mathbf e_2}$, with
$v_1,v_2$ being parameters of the  Painlev\'e III equation and
${\mathbf e_1}, {\mathbf e_2}$ being a canonical basis of 
${\mathbf V}$. Define next a symmetric
bilinear form  $\langle \cdot | \cdot \rangle$ in ${\mathbf V}$
such that $\langle {\mathbf e_i} | {\mathbf e_j} \rangle=\delta_{ij}$.
Then according to \cite{okap3} vectors
\begin{equation}
{\mathbf a_1}=  {\mathbf e_1} - {\mathbf e_2}, \;\;
{\mathbf a_2}=  {\mathbf e_2} 
\lab{B2roots}
\end{equation}
are the fundamental roots of the 
$B_2$ root system and
\begin{equation}
{\mathbf a_0}=  {\mathbf e_1} + {\mathbf e_2}
\lab{B2highroot}
\end{equation}
is its highest root.

Reflections in ${\mathbf V}$ with respect to 
the lines $\langle {\mathbf a_i} | {\mathbf v} \rangle= 0, i=1,2$
and $\langle {\mathbf a_0} | {\mathbf v} \rangle= -1$
generate transformations \rf{pi0vv}.
Geometrically, the transformations $s_2, \pi^2$ are reflections
in the hyperplane perpendicular to vectors ${\mathbf a_i},i=1,2$
and the transformation $\pi_0$
corresponds to reflections in the hyperplane $\{ {\mathbf v}: 
\langle {\mathbf a_0} | {\mathbf v} \rangle=-1\}$.
These hyperplanes are determined by 
the conditions $\langle {\mathbf a_1} | {\mathbf v} \rangle= 0,
\langle {\mathbf a_2} | {\mathbf v} \rangle= 0 $ and
$\langle {\mathbf a_0} | {\mathbf v} \rangle= -1$ or alternatively by
$v_1=v_2 , v_2=0$ and $ v_1=-1-v_2$, respectively.

The above discussion points to $B^{(1)}_2 \times \mathbb{Z}_2$ 
symmetry for $v_1,v_2,  \epsilon_0$ configuration space.
It is worthwhile to point out that thanks to the fact that we
extended the configuration space to include $\epsilon_0$ all our 
$B^{(1)}_2 $ B\"acklund transformations transform $z\to z$
without a need to include a $z\to -z$ transformation 
as in \cite{folding,forrester}. 


\subsection{Hamiltonian representations of PIII-PV system}
\label{subsection:Hamiltonian}

{In general case with 
$C,r_0,r_1$ taking arbitrary values it is useful 
to use Hamiltonian framework to study 
explicit form of symmetry operations 
of  equations \rf{big}.}

We start defining Hamiltonian representation  by 
introducing canonical variables $q,p$ through 
\begin{equation}
\begin{split}
f_1 &=   q\, ,  \qquad f_3=-q+r_1z^{-C} \\
f_2 &=-p \, , \qquad  f_0 = p +r_0 z^{C+\Omega} \, .
\lab{fpdict}
\end{split}
\end{equation}
Then equations \rf{big} can be summarized as two Hamilton equations:
\begin{equation}
\begin{split}
z q_z &= q \left(q-r_1 z^{-C}\right) \left(2p+z^{C+\Omega} \right)- 
\left(\alpha_1+\alpha_3+C \right) q\\
&+\alpha_1 r_1  z^{-C} +\epsilon_0 r_0 z^{C+\Omega}\\
z p_z &=  - p \left(p+ z^{C+\Omega} \right)
\left(2q-r_1z^{-C} \right) + 
(\alpha_1+\alpha_3+C) p \\&- 
\alpha_2  r_0 z^{C+\Omega} + 
\epsilon_1 r_1 z^{-C}
\end{split}
\lab{cceqs}
\end{equation}
obtained from the Hamiltonian:
\begin{equation}
\begin{split}
z H_{C} &= p  \left(p+r_0 z^{C+\Omega} \right) q  \left(q-r_1z^{-C} \right)-
\left(\alpha_1+\alpha_3+C\right) pq
\\&+\left(\alpha_1 r_1 z^{-C}  +
\epsilon_0 r_0 z^{C+\Omega} \right) p + 
\left( \alpha_2 r_0 z^{C+{\Omega}}+\epsilon_1 r_1 z^{-C} \right)  q
\end{split}
\lab{hamCC}
\end{equation}
{In the following two sub-subsections of this
subsection we will study
symmetries of the Hamiltonian system \rf{cceqs} for two possible
values of $C$: $C = -\Omega/2$  and  $C \ne -\Omega/2$.}

\subsubsection{The Case of $\mathbf{C =-\Omega/2}$}
\label{subsubsection:c=momega2}
{In this subsubsection we impose the condition 
\rf{Calphas} and study the second order differential equation of $q$. 
We will be able to show that all Hamilton equations 
for arbitrary $C$ such that $C =-\Omega/2$ can be derived
by simple rescaling from one particular Hamiltonian
structure associated with the fixed value of $C$ (taken here to be
$C=1/2$). This is important for evaluating a general nature of Dirac
reduction process shown in Section \ref{section:Dirac}.}

With  condition \rf{Calphas} holding the
relations \rf{condcbig} become \rf{condcbig1}.
In this context we define the canonical variables $q,p$ through
\begin{equation}
q = f_1 z^C, \qquad p = -f_2 z^{-C}
\lab{fpzdict}
\end{equation}
In this parametrization equations \rf{big} become
\begin{equation}
\begin{split}
z q_z &= q \left(q-r_1\right) \left(2p+r_0 z^{-2C}\right)- 
\left(\alpha_1+\alpha_3\right) q
+\alpha_1 r_1+\epsilon_0 r_0\\
z p_z &=  - p \left(p+r_0 z^{-2 C} \right)\left(2q-r_1\right) + 
(\alpha_1+\alpha_3) p - \alpha_2 r_0 z^{-2 C} - \epsilon_1 r_1 z^{-2 C}
\end{split}
\lab{Ceqsmots}
\end{equation}
Equations \rf{Ceqsmots} follow
from  the Hamiltonian \rf{hamC}:
\begin{equation}
\begin{split}
z H_C &= p  \left(p+r_0 z^{-2 C} \right) q  \left(q-r_1\right)-
\left(\alpha_1+\alpha_3\right) pq
+\left(\alpha_1 r_1+\epsilon_0 r_0\right) p 
\\&+ \left( \alpha_2 r_0  +\epsilon_1 r_1 \right) z^{-2 C} q
\end{split}
\lab{hamC}
\end{equation}
The canonical transformation $q \to q z^C, p \to pz^{-C}$
sends the Hamilton equations \rf{Ceqsmots} into the Hamilton
equations \rf{cceqs} for appropriate values of $C$.

Eliminating $p$ from the above equations \rf{Ceqsmots} and setting $r_0=1,
\epsilon_1=-1$   we obtain for $q_{zz}$ : 
\begin{equation} \begin{split}
q_{z_c z_c} & = - \frac{1}{z_c} q_{z_c}+ 
\left( \frac{1}{2q}+\frac{1}{2(q-r_1)} \right)
q_{z_c}^2 
- \frac{(\alpha_1+\alpha_3+2\alpha_2+2C-2s) q(q-r_1)}{4C^2 z_c } \\
&-\frac{ (\alpha_1+\alpha_3)(r_1(\alpha_3-\alpha_1)/2-\epsilon_0)q }
{4 C^2 z_c^2 (q-r_1)} 
- \frac{(\alpha_1r_1+\epsilon_0)^2}{8 C^2 z_c^2}
\left(\frac{1}{ q}+\frac{1}{q-r_1}
\right)\\
&+\frac{1}{8C^2} q (q-r_1) (2q-r_1)
\lab{seXP5}
\end{split}
\end{equation}
where $ z_c= z^{-2C}$.

Note that the term $\alpha_1+\alpha_3+2\alpha_2+2C$ can be rewritten 
$\alpha_2-\alpha_0$ for all the models with $C\ne0$.

It is interesting to note that the Hamilton equations \rf{Ceqsmots}
can be rewritten as 
\[ z_c\, p_{z_c} = -\frac{-1}{2C} \frac{\partial \left(z H_{C=-1/2}\right) }{\partial
q},\quad
 z_c\,  q_{z_c} = \frac{-1}{2C} \frac{\partial \left(z H_{C=-1/2}\right)  }{\partial
q}, \]
in terms of one single $H_{C=-1/2} $ Hamiltonian from equation 
\rf{hamChalf} {for $r_0=1,
\epsilon_1=-1$}. Thus the {special} case
{of} $C=-1/2$ carries
all information about {any} Hamiltonian system with 
{$C=-\Omega/2$}.
It is therefore worthwhile to now consider a special example of 
$C=-1/2$ with Hamiltonian $H_{C=-1/2} $  giving rise to equations 
{\rf{Hamceqs}} when we set $r_0=1$ and
$\epsilon_1=-1$.
Plugging $C=-1/2$ into \rf{seXP5} one obtains
 \begin{equation} \begin{split}
 q_{zz} & = - \frac1z q_z+ \left( \frac{1}{2q}+\frac{1}{2(q-r_1)} \right)
 q_z^2 
 - \frac{(\alpha_1+\alpha_3+2\alpha_2-1-2r_1) q(q-r_1)}{z } \\
 &-\frac{ (\alpha_1+\alpha_3)(r_1(\alpha_3-\alpha_1)/2-\epsilon_0)q }{z^2 (q-r_1)} 
 - \frac{(\alpha_1r_1+\epsilon_0)^2}{2z^2}\left(\frac{1}{ q}+\frac{1}{q-r_1}
 \right)\\
 &+\frac12 q (q-r_1) (2q-r_1)
 \lab{seP5}
 \end{split}
 \end{equation}
For $r_1=0$ the substitution:
\[
\tau=\sqrt{z}, \;\; q= \frac{y}{\sqrt{z}}
\]
into equation \rf{seP5} yields the Painlev\'e III equation:
\begin{equation}
y_{\tau \tau} = \frac{y_\tau^2}{y}-\frac{y_\tau}{\tau}-
\frac{4(\alpha_2-\alpha_0)}{\tau}y^2
-\frac{4\epsilon_0 (\alpha_1+\alpha_3)}{\tau}+2 y^3-\frac{4\epsilon_0^2}{y}
\lab{P3tau}
\end{equation}
We note that for $r_1\ne 0$ the parameter $\epsilon_0$
can be absorbed by simple 
redefinitions $\alpha_1 \to \alpha_1-\epsilon_0/r_1$,
 $\alpha_3 \to \alpha_3+\epsilon_0/r_1$,  this is contrary to the situation in
equation \rf{sP5q} with $C\ne-\Omega/2$, where the $\epsilon_0$-term could not be 
eliminated by redefinition of parameters.
Furthermore 
the simple change of transformation $q \to y=1-r_1/q$
casts equation \rf{seP5} in a standard Painlev\'e V form.
This is fully consistent with an observation that for $r_0\ne 0$ $ r_1
\ne0$ equation \rf{big} describes the $A^{(1)}_3$ Painlev\'e V system.

In summary, we have seen that the model for $C\ne0$ satisfying 
the condition \rf{Calphas}
effectively contains only one deformation parameter $r_1$ and the value
of $r_1$ determines the type of the Painlev\'e equation described by the
model with either the Painlev\'e III equation for $r_1=0$ and 
the Painlev\'e V equation for $r_1\ne 0$.

\subsubsection{The Case of $\mathbf{C \ne -\Omega/2} $}
\label{subsubssection:notomega}
{Since $C \ne -\Omega/2$ we can not automatically 
conclude here that 
the model is invariant under  $A^{(1)}_3$ extended affine Weyl group
for $r_0 \ne0, r_1\ne0$ and we have to involve the values of deformation parameters $\epsilon_i, i=0,1$
in the symmetry analysis.
}

Let us start with the simplest case of $\epsilon_1 =0$ and $\epsilon_0=0$.
In such case the Hamilton equations \rf{cceqs} of $H_{C}$ 
are invariant under transformations :
\[s_0(q)= q+\frac{\alpha_0}{ r_0 z^{C+\Omega}+p}, \; s_0(p)=p,\; 
s_0 (\alpha_{2 k})= -(-1)^k \alpha_{2k},\; 
s_0 (\alpha_{2k+1})= \alpha_{2k+1}+\alpha_0,
\]
$k=0,1$ as well as 
\[s_2(q)= q+\frac{\alpha_2}{p}, \; s_2(p)=p,\; 
s_2 (\alpha_{2 k})= (-1)^k \alpha_{2k},\; 
s_2 (\alpha_{2k+1})= \alpha_{2k+1}+\alpha_2,
\]
for $ k=0,1$.
In addition
as long as $\epsilon_0 =  0, \epsilon_1=0$  one can also define 
symmetry transformations
\[ \pi (q)= - p \frac{r_1z^{-C}}{r_0 z^{C+\Omega}} , 
\quad \pi (p)=\frac{r_0 z^{C+\Omega}}{r_1z^{-C}} 
\left(q- r_1z^{-C} \right),\quad
\pi(\alpha_i)= \alpha_{i+1}
\]
\[s_1(p) = p-\frac{\alpha_1}{q}, \;  s_1(q)=q,\;
s_1 (\alpha_{2k+1})= -(-1)^k \alpha_{2k+1},\;
s_1 (\alpha_{2k})= \alpha_{2k}+\alpha_1,
\]
for $k=0,1$ and $s_3 = \pi^2 s_1 \pi^2$.
Despite its unconventional form the above automorphism $\pi$ satisfies 
B\"acklund relations $\pi s_i=s_{i+1} \pi$ and together
with $s_i, i=0,1,2,3$ form the $A^{(1)}_3$ extended affine Weyl group
symmetry of the Hamilton equations \rf{cceqs} as long as $\epsilon_i=0, i=0,1$.
Note that due to $\epsilon_i=0, i=0,1$ it is in this case anyway 
possible to shift the
value of $C$ in \rf{big} to ensure that the condition \rf{Calphas}
is satisfied

If we set $\epsilon_1 =0$ and $\epsilon_0\ne0$,
the Hamilton equations \rf{cceqs} of $H_{C}$ 
are no longer invariant under $\pi, s_1,s_3$
but remain invariant under 
\[ \pi^2(q)= -q+r_1z^{-C}\; , 
\quad \pi^2 (p)=-p- r_0 z^{C+\Omega} ,\;
\pi^2(\alpha_i)= \alpha_{i+2},  \quad \pi^2 (\epsilon_0)=-\epsilon_0\]
as well as $s_0$ and $s_2 = \pi^2 s_0 \pi^2$.
Together $s_0,s_2, \pi^2$ 
form the extended affine Weyl group $A^{(1)}_1$.

For $\epsilon_0 \ne 0, \epsilon_1=0$ one can define birational symplectic transformations
\begin{align}
{\bar \pi}_0 (q) &= -\frac{\epsilon_0}{q}, \; {\bar \pi}_0 (p) = 
\frac{1}{\epsilon_0} \left( q^2p+\alpha_2q \right)\nonumber\\
{\bar \pi}_2 (q) &=  \frac{\epsilon_0}{q}, \; {\bar \pi}_2(p) =
\frac{-1}{\epsilon_0} \left( q^2(p+r_0 z^{C+\Omega})+(\alpha_1+\alpha_2+\alpha_3+C)q+
\epsilon_0 r_0 z^{C+\Omega} \right) .
\nonumber
\end{align}
that transform the Hamiltonian $H_{C}$ from
\rf{hamCC} into another polynomial in $p,q$  and thus 
preserve the holonomy of the system \cite{sasano}.

 For $r_1 = 0$ (and therefore $\epsilon_1$ effectively
eliminated from \rf{big})
these transformations give rise to new symmetries :
\begin{equation}\begin{split}
\pi_0(q)&= \frac{-\epsilon_0}{q}, \quad  
\pi_0 (p)= \frac{1}{\epsilon_0} \left( q^2p+\alpha_2 q \right)\\
\pi_0(\alpha_0)&= 2\Omega+2C -\alpha_0, \quad \pi_0(\alpha_2) = \alpha_2 \\
\pi_0(\alpha_1) &= \alpha_1+\alpha_0-\Omega-C, \quad  
\pi_0(\alpha_3) = \alpha_3+\alpha_0-\Omega-C 
\end{split}
\lab{pi02pq}
\end{equation}
and $\pi_2$ such that $\pi_2 =  \pi^2\pi_0\pi^2$ in which we 
recognize $B^{(1)}_{2}$ symmetries given in \rf{pi2}
in a different basis.

\subsubsection{The Case of $\mathbf{C=0}$}
\label{subsubsection:czero}
We now focus on the case of $C=0$ that  clearly falls into a 
category of $C \ne - \Omega/2$
since $\Omega$ is a constant different from zero.

All the equations and symmetry results of subsection 
\ref{subsection:Hamiltonian} and 
subsubsection \ref{subsubssection:notomega} follow 
just by inserting $C=0$ into the appropriate places.

Especially,  the Hamiltonian \rf{sasanohame} and equations \rf{eqsmots} 
are obtained from the Hamiltonian \rf{hamCC} and equations \rf{cceqs}
by setting  $C=0, r_0=1$ and $\epsilon_1=0$.
Eliminating $p$ from equations \rf{eqsmots} for $C=0$  yields
equation \rf{sP5q} :
\begin{equation} \begin{split}
q_{zz} & = - \frac1z q_z+ \left( \frac{1}{2q}+\frac{1}{2(q-r_1)} \right)
(q_z^2 -\epsilon_0^2)
- \frac{(\alpha r_1 -\epsilon_0 z \alpha_3) q}{z^2 (q-r_1)} \\
&-\frac{\beta r_1 (q-r_1)+\epsilon_0 r_1 z \alpha_1 }{z^2 q} 
- \frac{\gamma q (q-r_1) - \epsilon_0  (1+\alpha_1)}{z}
+\frac12 q (q-r_1) (2q-r_1)
\end{split}
\tag{1.5}
\end{equation}
with constants $\alpha, \beta, \gamma$ given by:
\begin{equation}
\alpha= \frac12 \alpha_3^2 ,\;\quad
\beta=-\frac12 \alpha_1^2 , \;\quad
\gamma= \alpha_2-\alpha_0\, .
\lab{abc}
\end{equation}
The extra parameters $r_1,\epsilon_0$ measure level of 
symmetry breaking.
\begin{itemize}
\item For $\epsilon_0=0, r_1\ne 0$ the above system is invariant under $s_i,
i=0,1,2,3, \pi$ and \rf{sP5q} becomes the Painlev\'e V equation for 
$y=(q-r_1)/q$ .

\item For $\epsilon_0 \ne 0, r_1= 0$ the above system is invariant under $s_i,
i=0,2, \pi^2, \pi_0,\pi_2$ and   \rf{sP5q} in this limit goes  to 
the Painlev\'e III equation.

\item For $\epsilon_0 \ne 0, r_1\ne 0$  the above Hamilton equations
are only invariant under $s_0, \pi^2$ and its composite $s_2$.
In view of the result establishing the Painlev\'e test for this
equation (see the end of this subsection below) this equation can be 
viewed  as a mixture of  Painlev\'e III and Painlev\'e  V equations.

\item
Finally, for the special case of $\epsilon_0 = 0, r_1= 0$ the above system has additional 
symmetries and can be solved by quadratures as discussed in
\ref{subsection:solvable}.
\end{itemize}
Let us now 
address the question whether equation
\rf{sP5q} passes the Painlev\'e test for arbitrary values
of deformation parameters $r_1, \epsilon_0$. 
First by inserting
\[ q(z)= a_0 (z-z_0)^\mu
\]
and  focusing on the dominant behavior near singularity
on both sides of equation
\rf{sP5q} we obtain 
\[ \mu(\mu-1) a_0 (z-z_0)^{\mu-2} = \frac{a_0^2 \mu^2  
(z-z_0)^{2\mu-2}}{ a_0 (z-z_0)^\mu}+a_0^3 (z-z_0)^{3 \mu} 
\]
with contributions on the right hand side originating from the second
and last term of the right hand side of  equation
\rf{sP5q}. 
This way we obtain: 
\[ a_0^2 =1 ,\qquad\quad \mu=-1 
\]
consistent with the Painlev\'e requirement that $\mu$ is a negative integer
for a movable pole with no branching.
Next to check the resonance condition  we plug
\[ q (z) = a_0 (z-z_0)^{-1} + \eta (z-z_0)^{-1+r}
\]
into  equation \rf{sP5q} and keep only the terms linear in $\eta$ to
obtain the resonance equation for $r$:
\[  (r+1)(r-2)=0
\]
which is identical to the resonance condition for PIII (see 
\cite{gromak}). 
This resonance structure suggests that a Laurent expansion
\begin{equation}
q (z) = \sum_{j=0}^{\infty} a_j  (z-z_0)^{j-1}=
a_0 (z-z_0)^{-1} + a_1 + h (z-z_0)+ a_3 (z-z_0)^{2} + a_4 (z-z_0)^{3}
+ \cdots
\lab{Laurent}
\end{equation}
expresses expansion around an arbitrary pole at $z_0$
where we identified $a_2=h$ as the single arbitrary coefficient.
Direct insertion yields 
\[ a_0^2=1, \qquad a_1 = \frac{1}{2 z_0} \left( z_0 r_1+
\alpha_2-\alpha_0 \pm 1 \right)
\]
and confirms that $a_2 = h$ is arbitrary with 
all higher coefficients $a_j, j\ge 3$
uniquely determined by lower coefficients of the Laurent expansion.
These observations verify successfully the Painlev\'e test for
equation \rf{sP5q}.

\subsection{ Solvable model: $r_1=0, \epsilon_0=0$ }
\label{subsection:solvable}
The case with both $r_1=0$ and $\epsilon_0=0$ is
solvable by quadratures. With $\epsilon_1$ effectively eliminated due
to  $r_1=0$ and $\epsilon_0$ put to zero the model becomes a submodel of PV. Introducing $F=f_1f_2$ one can rewrite
 in such case \rf{symP3C} as
\begin{subequations}\lab{Ff1}
\begin{align}
z^{-C+1}F_z &= f_1 \left(-F+\alpha_2 \right)
\lab{Ft}\\
z f_{1\, z}&=  -2 f_1 F+z^{-C}f_1^2+(C+\alpha_0+\alpha_2) f_1
\lab{f1t}
\end{align}
\end{subequations}
where we set  $r_0=1$ (in addition to $r_1=0$ and $\epsilon_0=0$).
One obtains from \rf{Ft}:
\begin{equation}
f_1 = \frac{z^{C+1}F_z}{\alpha_2-F}
\lab{f1fromF}
\end{equation}
Plugging this into \rf{f1t} yields 
\[
z F_{zz}= (z F_z-F)_z=\left(- F^2 + (-1+\alpha_0+\alpha_2)F \right)_z
\]
and we notice that dependence on $C$ dropped out in the above equation.
After integration the above Ricatti equation can be ``linearized'' 
by substituting:
\[
F= -z \left( \ln \psi \right)_z
\]
that yields the linear second order equation:
\[
z \psi_{zz} +(1-\alpha_0 -\alpha_2) \psi_z +K_1 \psi/z=0
\]
that appears to be a Bessel equation with solution
\begin{equation}
\psi(z)= A_1 \sqrt{z} J_{\sqrt{1-4K_1}} (2
\sqrt{1-\alpha_0-\alpha_2}\sqrt{z})
+A_2  \sqrt{z} 
Y_{\sqrt{1-4K_1}} (2 \sqrt{1-\alpha_0-\alpha_2}\sqrt{z})
\lab{bessel}
\end{equation}
where $K_1$ is an integration constant \cite{gromak-book}.

\section{Discussion}
\label{section:dicussion}

The $A^{(1)}_3$ Painlev\'e V equation is here modified  by 
introducing a new parameter and by addition of
two terms with deformation parameters.
Various symmetry structures emerging in the symmetric 
Painlev\'e III-V system obtained in such way are studied in the Hamiltonian 
formalism  and shown to depend on mutual relations
between deformation parameters, integration constants and the 
new explicit parameter of the underlying equations.   
The integrable origin of the symmetric  Painlev\'e III-V system 
is identified with the $4$-boson integrable hierarchy that reduces 
to the $A^{(1)}_4$ higher Painlev\'e equations in a self-similarity
limit.
The Dirac Lagrange multiplier reduction method is here applied to second-class
constraints of the $A^{(1)}_4$  Painlev\'e Hamiltonian system 
to obtain generic Hamiltonians of the Painlev\'e III-V system.

Various submodels of PIII-V model that  are 
governed by subgroups $A^{(1)}_3$, $B^{(1)}_2$ and $A^{(1)}_1$ 
are contained in a hybrid ordinary differential
equation that passes the Painlev\'e  test.
Due to this result this work relates to recent papers on mixing of Painlev\'e equations. 
A model that mixes the second and the third Painlev\'e equations
and possesses Painlev\'e property was obtained from a mixture of
mKdV-Sine Gordon integrable models \cite{kudryashov}.
Likewise a mixture of Lund-Regge and AKNS models was to shown to
reduce to the ordinary differential equation of Painlev\'e type \cite{ruy}.


The work is in progress on performing reduction of $6$-boson integrable model 
model to Painlev\'e VI and other models.

\vskip 0.5cm
{\bf  Acknowledgments}\\
Work of HA was performed when he was recipient of 2014-2015 Fulbright
U.S. Scholar Award. Support from the U.S.-Brazil Fulbright Commission
is gratefully acknowledged. AHZ and JFG were partially supported by
CNPq and DVR by Fapesp and Capes.

\end{document}